# Robust Extraction of Electron Energy Probability Function via Neural Network-Based Smoothing

June Young Kim

*Department of AI Semiconductor Engineering, Korea University, Sejong, South Korea*

**ABSTRACT**

Accurate determination of the electron energy probability function (EEPF) is vital for understanding electron kinetics and energy distributions in plasmas. However, interpreting Langmuir probe current-voltage (I–V) characteristics is often hindered by nonlinear sheath dynamics, plasma instabilities, and diagnostic noise. These factors introduce fluctuations and distortions, making second derivative calculations highly sensitive and error-prone. Traditional smoothing methods, such as the Savitzky-Golay (SG) filter and AC modulation techniques, rely on local data correlations and struggle to differentiate between noise and meaningful plasma behavior. In this study, we present a neural network-based machine learning approach for robust EEPF extraction, specifically designed to address the challenges posed by non-Maxwellian electron energy distributions. A multi-layer perceptron combined with ensemble averaging captures the global structure of the I–V characteristics, enabling adaptive and consistent smoothing without compromising physical fidelity. Compared to conventional SG filtering, the proposed method achieves superior smoothing of the second derivative, resulting in more stable and accurate EEPF reconstruction across the entire electron energy range. This capability confers a strong diagnostic advantage in beam-driven, low-pressure, or other non-equilibrium plasma conditions, where accurate characterization of non-Maxwellian EEPFs is essential.

E-mail: juneyoungkim@korea.ac.kr

**INTRODUCTION**

More than a century after Irving Langmuir coined the term "plasma" and introduced the Langmuir probe [1], this diagnostic technique remains a cornerstone in plasma research, particularly for extracting the electron energy probability function (EEPF). Despite its enduring relevance, the growing complexity of modern plasma environments highlights the need to strengthen this classical diagnostic tool with advanced data processing methods [2,3]. Traditional approaches, such as the Druyvesteyn method [4], derive the EEPF from the second derivative of the probe current with respect to the bias voltage. However, accurate interpretation of Langmuir probe current-voltage (I–V) characteristics is increasingly complicated by nonlinear phenomena such as sheath dynamics and plasma instabilities. These effects introduce intrinsic plasma fluctuations that distort probe measurements. Additionally, measurement noise inherent to diagnostic systems further amplifies uncertainty, making second derivative calculations particularly sensitive and prone to error. Existing smoothing techniques, including the Savitzky-Golay (SG) filter [5] and modulation-based approaches like AC superposition [6], focus on local data correlations within narrow regions of the I–V curve. While effective in reducing simple noise, these methods often fail to distinguish between random noise and meaningful physical variations. This limitation becomes especially problematic in the presence of highly nonlinear and irregular fluctuations, frequently resulting in distortion or loss of critical information in the reconstructed EEPF.

Over the past few decades, machine learning (ML) has emerged as a powerful tool for data analysis across various scientific and engineering disciplines. Since the introduction of the perceptron in the 1950s, ML algorithms—particularly neural networks—have evolved rapidly, driven by advances in computational power and algorithmic development [7]. Neural networks excel at capturing complex, nonlinear relationships in data and have demonstrated exceptional performance in denoising and smoothing tasks [8,9]. In plasma and nuclear fusion research, machine learning techniques are increasingly applied to diagnostic systems where data quality is critical. In Tokamak and Stellarator devices, ML-based approaches have been used to predict disruptions [10], and denoise diagnostic signals, enabling more accurate control of plasma stability and confinement [11,12]. For instance, neural network models have been employed to process signals from Thomson scattering and interferometry diagnostics, improving the measurement of electron temperature and density profiles even under severe fluctuation conditions [13−15]. These applications highlight the potential of ML not only to enhance data reliability but also to unlock deeper insights into plasma behavior.

Despite these advances, most applications of ML in plasma diagnostics have focused on complex or large-scale systems, while more conventional diagnostic methods have received relatively little attention. Here, we take a different approach by applying ML to Langmuir probe diagnostics—a well-established and widely used technique across various plasma regimes, ranging from low-temperature industrial plasmas to high-temperature fusion devices. To the best of our knowledge, this represents the first attempt to apply ML to Langmuir probe data analysis, with particular emphasis on the reliable extraction of the EEPF. We propose an ML-based approach using a multi-layer perceptron (MLP) combined with ensemble averaging to robustly extract the EEPF by capturing the global structure of the I–V characteristics.

**METHODS**

As shown in Fig. 1, the MLP is applied to smooth and functionally fit the I–V data obtained from Langmuir probe measurements. The MLP architecture consisted of two hidden layers, each containing 15 neurons. The number of hidden layers and neurons was determined empirically to balance model complexity and prevent overfitting. Each hidden layer utilized the hyperbolic tangent sigmoid activation function, allowing the network to capture nonlinear relationships between the probe voltage and current. The output layer employed a linear activation function to accommodate continuous regression. Network training was conducted using the Bayesian regularization backpropagation algorithm, which minimizes a combination of mean squared error and model complexity by introducing a weight decay term. The target training goal was set to $3.7 \times 10^{-6}$, corresponding to a prediction error of approximately 10 $\mu A$ for the probe current in the region below the plasma potential, extending to $-100$ V of the probe voltage. This range corresponds to approximately 100 eV of electron energy.

Prior to training, both the probe voltage $V_{measured}$ and measured current $I_{measured}$ were normalized using Z-score normalization to ensure numerical stability and improve training efficiency. The normalization was performed as $V_{norm} = V_{measured} - \mu_V/\sigma_V$ and $I_{norm} = I_{measured} - \mu_I/\sigma_I$, where $\mu_V$ and $\mu_I$ are the mean deviation, and $\sigma_V$ and $\sigma_I$ are the standard deviations of the measured voltage and current, respectively. This normalization step ensures that both input and output data are scaled to have zero mean and unit variance, allowing more efficient training of the MLP model.

Upon completion of the training phase, a dense voltage grid $V_{dense}$ was generated, covering the same voltage range with a fixed step size (e.g., 0.01V). This dense grid was normalized using the same Z-score parameters and subsequently fed into the trained MLP models to predict the corresponding current values. The outputs from the multiple MLP models were averaged to obtain the ensemble prediction, which was then denormalized.

To enhance robustness and reduce sensitivity to random weight initialization, an ensemble learning strategy was adopted. Multiple independent MLP models were trained with different random initial weights, and their outputs were averaged to produce the final predicted I–V curve. The ensemble-averaged normalized current was computed as: $I_{ensemble} = 1/N_m \sum_{i=1}^{N} I_{MLP,i}$, where $N_m$ is the number of ensemble models and $I_{MLP,i}$ is the predicted normalized current from the $i^{th}$ MLP model for the normalized $V_{dense}$. Finally, the ensemble-averaged current was denormalized to yield the final smoothed I–V characteristic using the inverse Z-score transformation.

The resulting smoothed I–V data was then processed to obtain the first and second derivatives with respect to the probe voltage. The second derivative of the I-V curve $I_e''(\varepsilon)$ is proportional to EEPF $f_e(\varepsilon)$ as follows [2]: $f_e(\varepsilon) = 2m_e/(e^2 A)(2/m_e)^{1/2} I_e''(\varepsilon)$, where $m_e$, $e$, $\varepsilon$, and $A$ are the electron mass, elementary charge, electron energy in volts, and probe area, respectively. The electron density $n_e$ and effective electron temperature $T_e$ corresponding to the mean electron energy are determined by integrating the EEPF as follows: $n_e(\varepsilon) = \int_0^\infty \varepsilon^{1/2} f_e(\varepsilon) d\varepsilon$, and $T_e = 2/3 \int_0^\infty \varepsilon^{3/2} f_e(\varepsilon) d\varepsilon / \int_0^\infty \varepsilon^{1/2} f_e(\varepsilon) d\varepsilon$. The plasma potential $V_p$ was identified as the zero-crossing point of the second derivative curve.

A single I–V characteristic, measured at the radial center of the plasma, was used as the dataset for training and evaluating the proposed MLP model. The experiments were performed in a Penning source at a magnetic field strength $B_z$ of 147 G or varying $B_z$ at argon pressure of 0.46 mTorr. The cathode voltage was fixed at $-60$ V, and the current was 3.0 A within 10% of the variation. Details of the apparatus configuration can be found in previous work [16]. Under these conditions, the plasma is considered collision-less at the relevant length scale (i.e., probe tip radius of 0.15 mm and the Debye length of 0.01 mm must be less than the electron-neutral collisional mean free path ($\sim$1 m) at 0.46 mTorr).

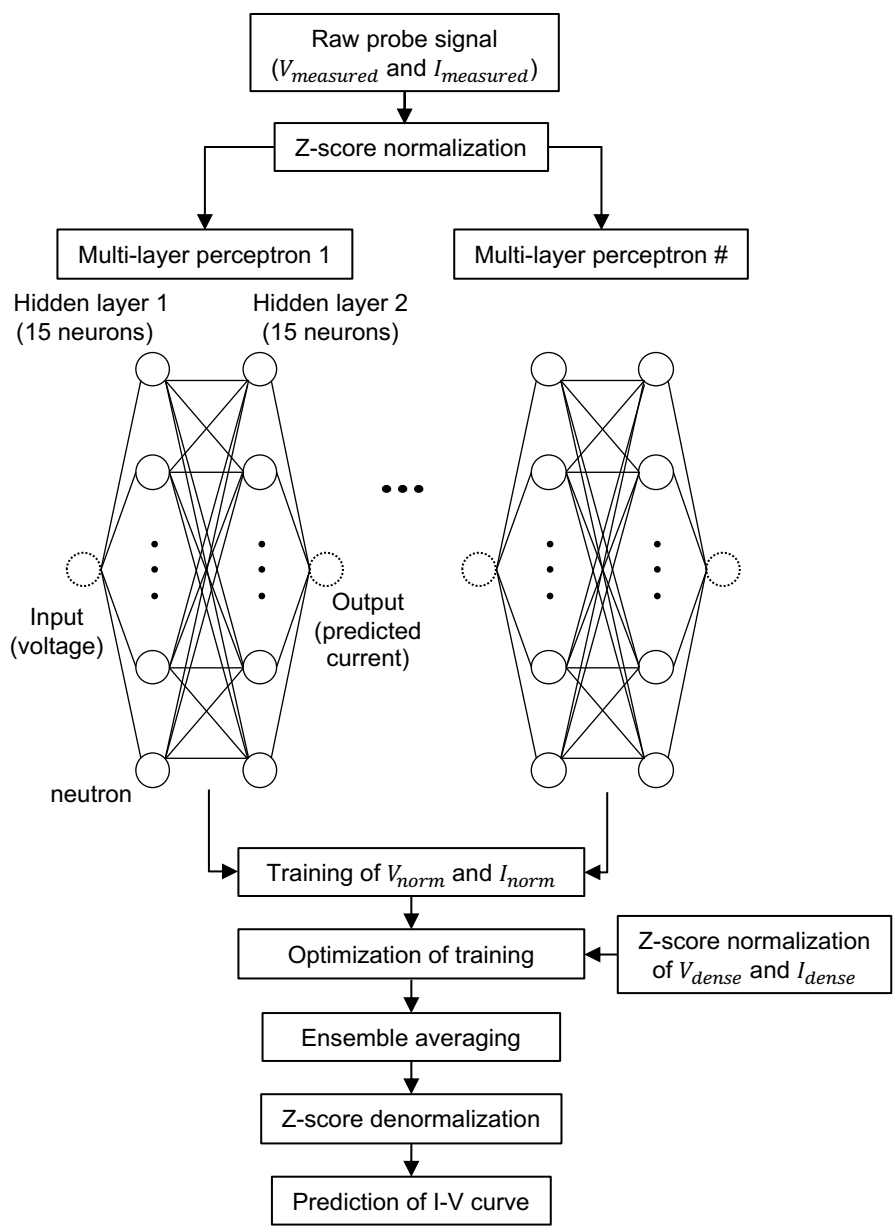

Figure 1. Procedural diagram of I–V Curve processing using MLP.

**RESULTS AND DISCUSSION**

Figures 2(a–c) compare the EEPFs derived from the raw I–V characteristics, the SG filter, and the proposed MLP-based method. In Fig. 2(a), the raw second derivative yields EEPFs with substantial scattering. The SG filter with a small window size (11 points) improves smoothing in the low-energy domain but remains ineffective at suppressing fluctuations at higher energies. By contrast, the MLP-based approach produces EEPFs that are smoothly connected across the entire energy range with minimal fluctuation. As the window size increases [Fig. 2(b)], high-energy fluctuations are progressively reduced; however, this comes at the cost of physical distortion. Notably, the low-energy peak of the EEPF shifts to higher energies. In particular, for non-Maxwellian EEPFs exhibiting two distinct slopes, the SG filter introduces significant distortion near the transition region between the two electron populations (around 25 eV), where accurate curvature preservation is critical. This breakdown arises from the fixed-window polynomial fitting, which inevitably spans both populations and fails to accommodate the abrupt change in slope—highlighting a fundamental limitation of the method. In contrast, the MLP model [Fig. 2(c)] maintains both the smoothness and the physical fidelity of the EEPF across all energy ranges. Increasing the ensemble size improves overall stability, but even a single trained network is sufficient to capture the essential features without spurious oscillations. The MLP's data-adaptive smoothing provides a more reliable framework for interpreting non-Maxwellian electron kinetics.

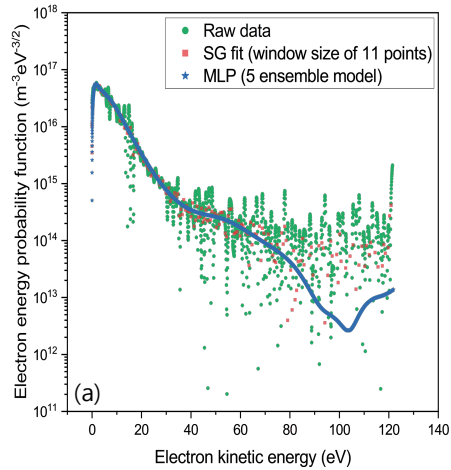

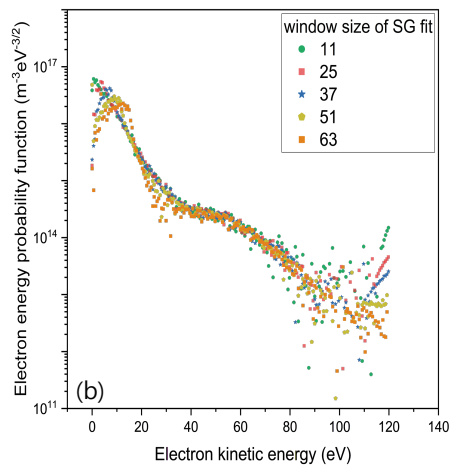

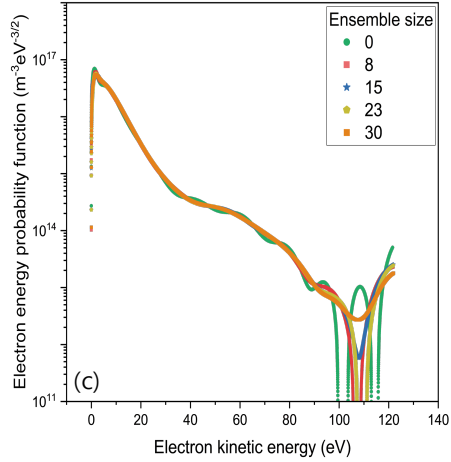

Figure 2. (a) EEPFs obtained from the second derivative of raw data (green), SG filter with a window size of 11 (red), and MLP model with 5 ensemble networks (blue). (b) EEPFs derived using the SG filter with various window sizes. (c) EEPFs predicted by the proposed MLP-based method with different ensemble sizes.

To quantitatively assess how smoothing methods influence the extracted plasma parameters, the plasma potential $V_p$, $T_e$, and $n_e$ were computed across a range of smoothing configurations. Figures 3(a–c) show the trends of these parameters with respect to the SG filter window size. While moderate window sizes offer stable parameter estimation, larger windows cause significant deviations—particularly for $T_e$ and $n_e$—due to over-smoothing and distortion in the second derivative of the I–V curve. This is especially problematic near the plasma potential, where fine variations critically affect the accuracy of $T_e$ and $n_e$. In contrast, Figures (d–f) present the same plasma parameters derived using the MLP-based approach with increasing ensemble size. All three parameters show remarkable stability regardless of the number of ensemble models, demonstrating the robustness of the neural network model to initialization randomness. Notably, even a single model without ensemble averaging yields accuracy comparable to that of larger ensembles, reinforcing the reliability and computational efficiency of the MLP-based method for plasma parameter inference.

To assess the reproducibility of the proposed MLP-based EEPF reconstruction, 30 repeated trials were performed under identical input conditions for two ensemble configurations: a single model and five independent models (ensemble size = 5). As shown in Fig. 4(a–c), the extracted plasma parameters — $V_p$, $n_e$, and $T_e$— exhibited minimal variation across trials in both cases. However, quantitative analysis reveals that ensemble averaging notably improves repeatability. Specifically, the standard deviations decreased from 0.249 V to 0.143 V for $V_p$, from $3.17 \times 10^{16}$ m$^{-3}$ to $1.66 \times 10^{16}$ m$^{-3}$ for $n_e$, and from 0.122 eV to 0.067 eV for $T_e$. These improvements confirm that the ensemble strategy effectively mitigates the influence of random weight initialization and enhances the stability of the machine learning output without compromising accuracy.

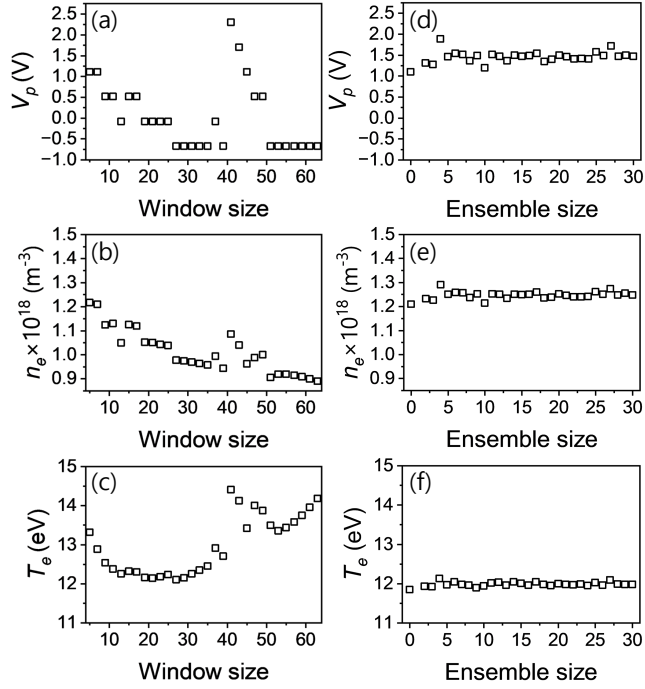

Figure 3. Comparison of plasma parameters obtained using different smoothing techniques. (a–c) $V_p$, $n_e$, and $T_e$ extracted using SG filters with varying window sizes. (d–f) Same plasma parameters derived using the MLP-based ensemble method with varying ensemble sizes.

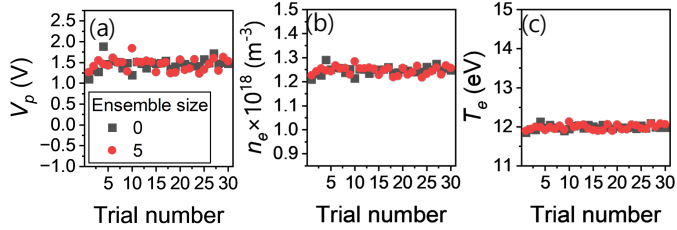

Figure 4. Reproducibility analysis of plasma parameters extracted by the MLP-based method across 30 repeated trials. (a) $V_p$, (b) $n_e$, and (c) and $T_e$ shown for two cases: without ensemble averaging (black rectangular) and with five-model ensemble averaging (red circles).\

To quantitatively evaluate the smoothness of the extracted EEPF, two metrics were employed: the total variation and the high-frequency (HF) power spectral density (PSD). The total variation quantifies the overall gradient fluctuation and is defined as $\sum_{i=1}^{N-1}|f_e(\varepsilon_{i+1}) - f_e(\varepsilon_i)|$, where $N$ is the total number of data points. The HF power was estimated using the periodogram method, and it captures the spectral energy in the upper half of the Nyquist frequency domain: $HF = \int_{\nu_c}^{\nu_{max}} PSD(\nu)d\nu$, where $\nu_c$ is the cutoff frequency at half of the maximum frequency $\nu_{max}$. For the voltage interpolation step of 0.01 V, the effective sampling frequency is 100 Hz. Consequently, $\nu_{max}$ is 50 Hz, and $\nu_c$ for high-frequency integration is set at 25 Hz.

Figures 5(a) and 5(b) show the variation of total variation and HF power, respectively, for the EEPFs obtained using SG filtering as the window size is varied. Both metrics exhibit strong sensitivity to the choice of window size. In contrast, Figs. 5(c) and 5(d) present the same metrics for the EEPFs reconstructed using the proposed MLP-based ensemble method as a function of ensemble size. Notably, both the total variation and HF power remain relatively stable and low across different ensemble sizes, indicating that the neural network model provides consistently smooth results. This suggests that the machine learning approach can eliminate high-frequency noise more robustly than traditional local polynomial filtering methods.

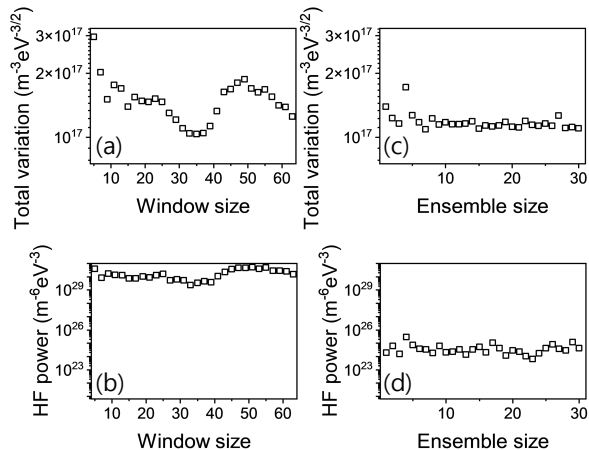

Figure 5. Quantitative evaluation of EEPF smoothness. (a) Total variation and (b) HF power of the EEPF obtained via SG filtering as the window size varies. (c) Total variation and (d) HF power of the EEPF obtained via the proposed MLP ensemble model as a function of ensemble size.

To evaluate whether a fixed MLP configuration can reliably reproduce physically meaningful plasma behavior, we applied the proposed machine learning model—comprising two hidden layers with 15 neurons each and an ensemble size of 2—to Langmuir probe measurements taken under varying $B_z$ in a beam-generated source [16]. The aim was to validate the robustness and generalizability of the method without tuning model parameters for each condition. Figures 6 presents the resulting plasma parameters and EEPFs. As shown in panels (a–c), the ML approach (red circles) yields smoother and more consistent trends compared to the conventional SG method (black squares). Specifically, $V_p$ and $T_e$ decreases monotonically with $B_z$, while $n_e$ exhibits a clear upward trend. The characteristic decrease in the potential is commonly observed when a magnetic field is applied perpendicular to the cathode surface and is associated with enhanced axial confinement in magnetized plasmas [16]. Such magnetic confinement also promotes increased ionization and suppresses radial diffusion, leading to a significant rise in $n_e$ as $B_z$ increases. The EEPFs shown in panel (d) further illustrate the superior robustness of the ML-based approach. The SG-derived EEPFs (scatter symbols) exhibit significant fluctuation across the entire energy range, particularly at high energies where numerical differentiation becomes unstable. In contrast, the ML method (solid-line appearance due to dense energy sampling) produces smooth, physically plausible distributions, even in regimes where non-Maxwellian features are present. The ML-based EEPFs also clarify the observed decrease in $T_e$ with increasing $B_z$, as the population of high-energy electrons progressively diminishes due to improved confinement and collisional thermalization.

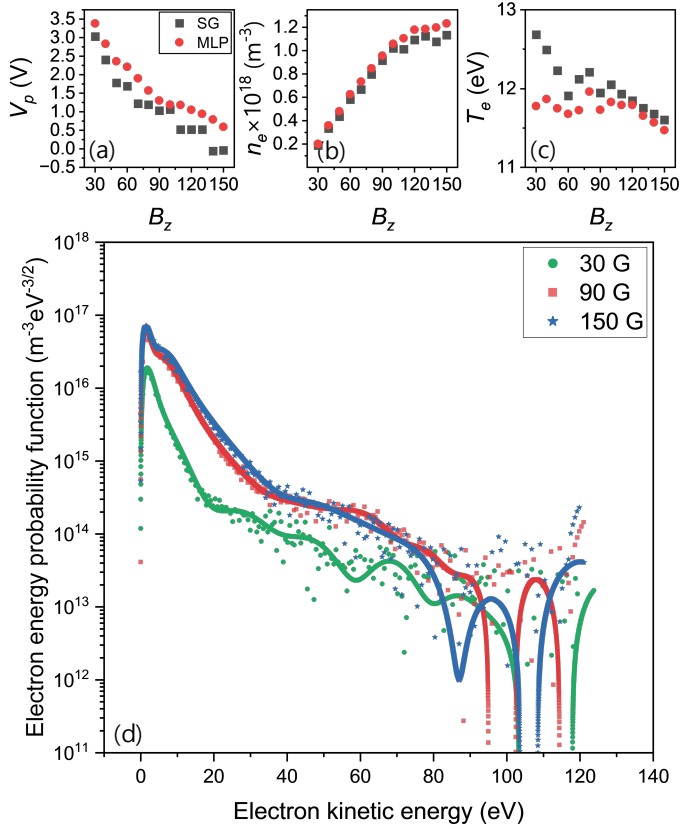

Figure 6. Comparison of plasma parameters and EEPFs obtained under various $B_z$. (a) $V_p$, (b) $n_e$, and (c) $T_e$ as functions of $B_z$. (d) Corresponding EEPFs at selected magnetic fields (30, 90, and 150 G). SG results are plotted as scattered points, while ML outputs appear as smooth lines due to dense energy sampling.

**CONCLUSIONS**

In this study, a neural network-based smoothing technique was introduced for the accurate and robust extraction of the EEPF from Langmuir probe diagnostics. By employing a MLP architecture in conjunction with ensemble averaging, the proposed approach effectively mitigates the influence of plasma fluctuations and diagnostic noise, which often compromise the reliability of conventional derivative-based methods. Compared to the widely used SG filter, the MLP-based method exhibited superior performance in both qualitative and quantitative evaluations. The ensemble-averaged model yielded stable reconstructions across the entire energy domain, successfully preserving non-Maxwellian features that are frequently distorted or lost under fixed-window polynomial smoothing.

The robustness of the MLP approach was further confirmed through its low sensitivity to ensemble size and strong reproducibility across repeated trials. Moreover, the plasma parameters extracted from the MLP-smoothed I–V curves demonstrated physically consistent trends with increasing magnetic field strength—namely, a monotonic change in plasma potential and electron density, accompanied by a slight decrease in electron temperature—reflecting enhanced magnetic confinement and reduced electron losses. Since accurate extraction of the electron velocity distribution function relies critically on precise ion current curve fitting from the I–V characteristics, the proposed method is expected to significantly enhance the fidelity of this process by providing a clean and physically consistent baseline.

ACKNOWLEDGEMENT

This research was supported by the National Research Foundation of Korea (NRF) grant funded by the Korea Government (MSIT) (RS-2023-00208968).

AUTHOR DECLARATIONS
**Conflict of Interest**
The authors have no conflicts to disclose.

DATA AVAILABILITY

The data that support the findings of this study are available within the article.

REFERENCES

[1] H. M. Mott-Smith and I. Langmuir, "The theory of collectors in gaseous discharges," Phys. Rev. 28(4), 727–763 (1926).
[2] F. F. Chen, "Langmuir probe diagnostics," in IEEE–International Conference on Plasma Science (ICOPS), Jeju, Korea (IEEE, 2003).
[3] R. B. Lobbia and B. E. Beal, "Recommended practice for use of Langmuir probes in electric propulsion testing," J. Propuls. Power 33, 566 (2017).
[4] M. J. Druyvesteyn, "Der niedervoltbogen," Z. Phys. 64(11), 781–798 (1930).
[5] A. Savitzky and M. J. E. Golay, "Smoothing and differentiation of data by simplified least squares procedures," Anal. Chem. 36, 1627–1639 (1964).
[6] J. L. Jauberteau and I. Jauberteau, "Determination of the electron energy distribution function in the plasma by means of numerical simulations of multiple harmonic components on a Langmuir probe characteristic—measurements in expanding microwave plasma," Meas. Sci. Technol. 18, 1235 (2007).
[7] F. Murtagh, "Multilayer perceptrons for classification and regression," Neurocomputing 2(5), 183–197 (1991). https://doi.org/10.1016/0925-2312(91)90023-5
[8] K. Noda, Y. Yamaguchi, K. Nakadai, H. G. Okuno, and T. Ogata, "Audio-visual speech recognition using deep learning," Appl. Intell. 42, 722 (2015).
[9] Y. Jia and J. Ma, "What can machine learning do for seismic data processing? An interpolation application," Geophysics 82.3 (2017): V163-V177.
[10] A. Piccione, J. W. Berkery, S. A. Sabbagh, and Y. Andreopoulos, Nucl. Fusion 60, 046033 (2020).
[11] G. Hu, T. Zhou, and Q. Liu, "Data-driven machine learning for fault detection and diagnosis in nuclear power plants: A review," Front. Energy Res. 9, 663296 (2021).
[12] J. Seo, S. Kim, A. Jalalvand, R. Conlin, A. Rothstein, J. Abbate, K. Erickson, J. Wai, R. Shousha, and E. Kolemen, "Avoiding fusion plasma tearing instability with deep reinforcement learning," Nature 626(8000), 746–751 (2024).
[13] R. Fischer, A. Dinklage, and E. Pasch, "Bayesian modelling of fusion diagnostics," Plasma Phys. Control. Fusion 45(7), 1095–1111 (2003).
[14] S. Kwak, J. Svensson, S. Bozhenkov, J. Flanagan, M. Kempenaars, A. Boboc, Y.-C. Ghim, and JET Contributors, "Bayesian modelling of Thomson scattering and multichannel interferometer diagnostics using Gaussian processes," Nucl. Fusion 60(4), 046009 (2020).
[15] A. Pavone, A. Merlo, S. Kwak, and J. Svensson, "Machine learning and Bayesian inference in nuclear fusion research: an overview," Plasma Phys. Control. Fusion 65, 053001 (2023).
[16] J. Y. Kim, J. Y. Jang, J. Choi, J. I. Wang, W. I. Jeong, M. A. I. Elgarhy, G. Go, K. J. Chung, and Y. S. Hwang, "Magnetic confinement and instability in partially magnetized plasma," Plasma Sources Sci. Technol. 30(2), 025011 (2021).